\documentclass[aps,preprint,groupedaddress]{revtex4}
\usepackage{graphicx}

\begin{document}
\title{Quantum Mechanics and the Time Travel Paradox}
\author{David T. Pegg}
\affiliation {School of Science, Griffith University, Nathan, Brisbane, Q 4111, Australia}
\date{\today}

\begin{abstract}
The closed causal chains arising from backward time travel do not lead to
paradoxes if they are self consistent. This raises the question as to how
physics ensures that only self-consistent loops are possible. We show that,
for one particular case at least, the condition of self consistency is
ensured by the interference of quantum mechanical amplitudes associated with
the loop. If this can be applied to all loops then we have a mechanism by
which inconsistent loops eliminate themselves.
\end{abstract}

\maketitle

\section{INTRODUCTION}

This paper is based on a talk given at a conference in Naples 
\cite{Nap}. It is well known that if backward time travel could be implemented, or if
the present could shape the past by some other means, then closed causal
chains, or causal loops, could be formed. The possibility of causal loops
must arise because changing, or rewriting, the past is not particularly
meaningful. If these loops are not self consistent, paradoxes arise \cite{Nov}. 
The best known of these is the grandfather paradox: a man travels
into the past to kill his grandfather before his grandfather meets his
grandmother. If he succeeds then we have the paradox that the man is never
born, thus the grandfather is not killed, thus the man is born and so on. A
simpler and more direct example is autoinfanticide, whereby a man travels
into the past and shoots himself as a baby. A different example is where a
man travels back and introduces his parents to each other \cite{Hor}. Here
we have a paradox if he fails in his task. A more direct version of this
latter type of paradox involves a robot, let us say, which enters a time
machine in the present. When it emerges from the machine in the past it is
sent through a rejuvenator which removes its memory and programs it to walk
around until it enters the time machine in the present. A paradox occurs if
something prevents the robot from entering the time machine in the present.

Sometimes the possibility of such paradoxes is taken as an argument against
the possibility of backward time travel. It can, however, be argued that,
because of continuity in nature, self-consistent loops or cycles in these
situations always exist and it is only these cycles that nature allows \cite{Sch}. 
Somehow the physics sorts things out so that the probability of
inconsistent cycles is zero. We shall refer to this argument as the
principle of self consistency \cite{Ear}. An example of a self-consistent
solution to the autoinfanticide cycle is the scenario in which the man
raises his rifle to shoot himself as a baby but, because of his bad
shoulder, misses the baby's heart and hits it in the shoulder \cite{Feyn}.

If we accept that self-consistent loops are reasonable, the question arises
as to the mechanism by which physics disallows inconsistent loops and
ensures self consistency. In this paper we show that quantum mechanics may
provide an answer to this question.

\section{QUANTUM MECHANICS}

Quantum mechanics is essentially concerned with the preparation of a system
and its subsequent measurement. The measurement device can be described
mathematically by a probability operator measure (POM) \cite{He}. Each
element $\hat{\Pi}_j$ of the POM corresponds to a possible outcome event 
{\it j} of the measurement. All the POM elements are non-negative definite,
that is, their expectation values for any state of the system are greater
than or equal to zero. The POM elements sum to the unit operator, ensuring
that the probability that there is some outcome of the measurement is unity.
The POM elements themselves indicate the states the system can be ``found
in''. The preparation device can be described mathematically by a collection
of preparation device operators (PDO) $\hat{\Lambda}_i$ associated with
possible preparation outcomes \cite{At}. These non-negative definite
operators indicate the possible states the system can be ``prepared in''.
The {\it a priori }probability $P(i)$ for a preparation event {\it i }is
given by the trace Tr$\hat{\Lambda}_i$. The trace of the sum of all the 
$\hat{\Lambda}_i$ is unity, ensuring that the probability that there is some
outcome of the preparation is unity. An essential feature of quantum
mechanics, as opposed to classical mechanics, is that, even if there is no
evolution of the system between the preparation and measurement events, the
state the system is found in is not necessarily the state the system is
prepared in.

\subsection{Predictive formalism}

The usual formalism of quantum mechanics is predictive. Between preparation
and measurement we assign to the system a state on the basis of our
knowledge of the preparation event. If we have no knowledge of the outcome
of the preparation, the best we can do is to assign a density operator
immediately after preparation equal to the sum of the PDO's. If, however, we
know the outcome was the preparation event {\it i}, we assign the normalized
density operator $\hat{\rho}_i^{pred}=$ $\hat{\Lambda}_i/$Tr$\hat{\Lambda}_i$
immediately after preparation. This state evolves unitarily forward in time
to become $\hat{U}\hat{\rho}_i^{pred}\hat{U}^{\dagger}$ at the time of
measurement. $\hat{U}$ is the usual forward time unitary evolution operator.
Upon measurement there is a discontinuous change, or collapse, of the state
to that corresponding to the POM element associated with the particular
measurement outcome {\it j}. The probability for the measurement outcome 
{\it j,} given the preparation outcome {\it i,} is given by 
\begin{equation}
P(j|i)=\text{Tr}[\hat{U}\hat{\rho}_i^{pred}\hat{U}^{\dagger }\hat{\Pi}_j].
\label{e1}
\end{equation}

As the name indicates, the predictive formalism is particularly useful for
predicting outcomes of measurement events based on knowledge of preparation
events. We can use the formalism, in conjunction with Bayes' theorem, to
retrodict outcomes of preparation events given measurement events\cite{Bay}.
From Bayes' theorem we have 
\begin{equation}
P(i|j)=\frac{P(j|i)P(i)}{P(j)}=\frac{P(j|i)P(i)}{\sum_iP(j|i)P(i)}.
\label{e2}
\end{equation}

When the known outcome {\it i} of the preparation device corresponds to a
pure state $\left| p_i\right\rangle$, both $\hat{\Lambda}_i$ and 
$\hat{\Lambda}_i/$Tr$\hat{\Lambda}_i$ are just 
$\left| p_i\right\rangle\left\langle p_i\right|$. Also when the known outcome {\it j }of the
measurement device corresponds to a pure state $\left| m_j\right\rangle$
both $\hat{\Pi}_j$ and $\hat{\Pi}_j/$Tr$\hat{\Pi}_j$ are just 
$\left|m_j\right\rangle \left\langle m_j\right|$. It is simpler in such cases to
work with the associated pure states instead of the PDO's and POM elements.
Expression (\ref{e1}) then becomes simply $\left| \left\langle m_j\right| 
\hat{U}\left| p_i\right\rangle \right| ^2$.

An interesting situation arises when we have prepared two systems {\it a }
and {\it b }in an entangled pure state, which evolves to, say, 
\[2^{-1/2}(\left| a_1\right\rangle _a\left| b_1\right\rangle _b+\left|
a_2\right\rangle _a\left| b_2\right\rangle _b) \]
at the time of the measurement of system {\it b}, which is measured before
system {\it a}. If the measurement outcome is the POM element corresponding
to state $\left| b_1\right\rangle_b$, then, in the predictive formalism,
the measurement projects this state onto the entangled state, giving the
state $\left| a_1\right\rangle _a$ after normalization. This collapsed state
continues to evolve forwards in time until system {\it a} is measured.

\subsection{Retrodictive formalism}

A less usual, but equally valid, quantum mechanical formalism is the
retrodictive formalism \cite{Ah}. Here the state of a system between
preparation and measurement is assigned on the basis of our knowledge of the 
{\it measurement} event. If we know the outcome was the measurement event 
{\it j}, we assign the normalized density operator $\hat{\rho}_j^{retr}=\hat{
\Pi}_j/$Tr$\hat{\Pi}_j$ to the system immediately before measurement \cite
{Bay}. This state evolves unitarily {\it backwards} in time to become $\hat{U
}^{\dagger }\hat{\rho}_j^{retr}\hat{U}$ at the time of preparation. Upon
preparation there is a discontinuous change, or collapse, of the state to
that corresponding to the PDO associated with the particular preparation
outcome {\it i}. The probability for the preparation outcome {\it i} given
the measurement outcome {\it j} is given by 
\begin{equation}
P(i|j)=\frac{\text{Tr}(\hat{U}^{\dagger }\hat{\rho}_j^{retr}\hat{U}\hat{
\Lambda}_i)}{\sum_i[\text{Tr}(\hat{U}^{\dagger }\hat{\rho}_j^{retr}\hat{U}
\hat{\Lambda}_i)]}.  \label{e4}
\end{equation}
The denominator ensures the probabilities for all possible preparation
events sum to unity. It is not difficult to show, using the cyclic property
of the trace, that this is the same as the formula for $P(i|j)$ in (\ref{e2}
) obtained from the predictive formalism plus Bayes' theorem. Use of the
retrodictive formalism for this purpose is more direct, however, which is
why it is becoming useful for solving the basic quantum communication
problem \cite{Amp}, in which Bob measures the state of a quantum system sent
by Alice and has to retrodict the state she selected to send. Our interest
now, however, is the application of the retrodictive formalism to entangled
systems.

Suppose the result of a measurement on two systems {\it a }and {\it b }%
corresponds to the POM element associated with the pure {\it retrodictive}
entangled state which evolves backwards in time to

\[
2^{-1/2}(\left| a_1\right\rangle _a\left| b_1\right\rangle _b+\left|
a_2\right\rangle _a\left| b_2\right\rangle _b) 
\]
at the time of the preparation of system {\it b}, which is prepared after
system {\it a} is prepared. If the preparation outcome is the PDO
corresponding to state $\left| b_1\right\rangle _b$, then we project this
state onto the entangled state, renormalize, and obtain the state $\left|
a_1\right\rangle _a.$ This collapsed retrodictive state continues to evolve
backwards in time until the time of the preparation of system {\it a}.

An even more interesting case arises when a retrodictive state from a
measurement of system {\it c} at time $t_m$ evolves backwards to a state $%
\left| c_1\right\rangle _c$ at the time $t_p$ of a preparation outcome that
corresponds to the entangled state

\[
2^{-1/2}(\left| c_1\right\rangle _c\left| d_1\right\rangle _d+\left|
c_2\right\rangle _c\left| d_2\right\rangle _d). 
\]
The projection of $\left| c_1\right\rangle _c$ onto this state collapses it
to $\left| d_1\right\rangle _d$, which can be associated with a predictive
state evolving forwards in time from $t_p$.

\section{BEAM SPLITTERS}

A method of making predictive entangled states is to prepare the light in
the input modes {\it b }and {\it c }of a beam splitter in the vacuum and
one-photon states with preparation devices P0 and P1 respectively. We can
then write the predictive input state as $\left| 0\right\rangle _b\left|
1\right\rangle _c$. To be specific, let the beam splitter, as shown in Fig.
1(a), be symmetric and reflect as much light as it transmits. The action of
the beam splitter can be represented by a unitary operator $\hat{R}$. It is
possible to show that for this 50/50 beam splitter, the action of the beam
splitter on the input state is to transform it to the entangled state \cite
{Book} 
\begin{equation}
\hat{R}\left| 0\right\rangle _b\left| 1\right\rangle _c=2^{-1/2}(\left|
0\right\rangle _b\left| 1\right\rangle _c+i\left| 1\right\rangle _b\left|
0\right\rangle _c)  \label{e5}
\end{equation}
in the output of the beam splitter. This shows that there is an equal chance
for the photon to remain in mode {\it c}, that is to be transmitted, and for
it to be reflected into mode {\it b}. It is sometimes convenient to regard
P0, P1 and the beam splitter as a combined preparation device that generates
the predictive state $2^{-1/2}(\left| 0\right\rangle _b\left| 1\right\rangle
_c+i\left| 1\right\rangle _b\left| 0\right\rangle _c)$ if the outcomes of P0
and P1 correspond to zero and one photons. This entangled state then
propagates forward in time. If, at a later time $t_m$, a measurement on the
field in mode {\it c }shows that the photon is in mode {\it c} then,
according to the usual description of the measurement process in the
predictive formalism, the field instantaneously collapses at $t_m$ to $
_c\left\langle 1\right| (\left| 0\right\rangle _b\left| 1\right\rangle
_c+i\left| 1\right\rangle _b\left| 0\right\rangle _c)=\left| 0\right\rangle
_b$ after normalization. That is, the field in mode {\it b }changes suddenly
to the vacuum simultaneously with the measurement of the field in mode {\it 
c, }even though this measurement event can take place an appreciable
distance away. This, of course, is just an illustration of the well-known
Einstein-Podolsky-Rosen paradox involving instantaneous collapse over a
large distance. The vacuum state $\left| 0\right\rangle _b$, being a
predictive state, continues to propagate forwards in time.

\begin{figure}
\includegraphics[width=0.60\textwidth]{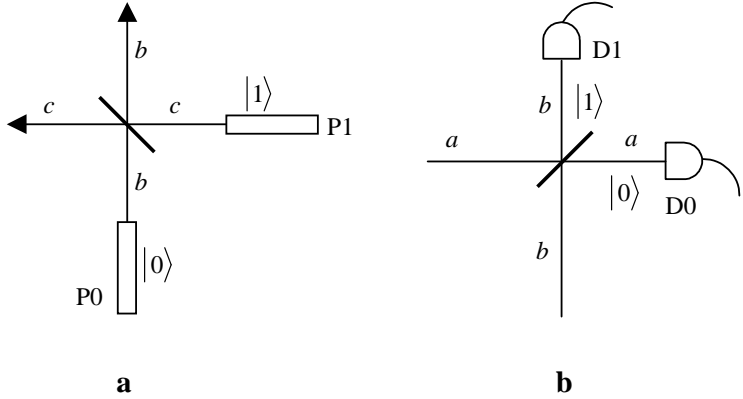}
\caption{Beam splitters. In (a) preparation devices P0 and P1 prepare
vacuum and one-photon states in input modes $b$ and $c$. In (b)
photodetectors D1 and D0 detect one and zero photons in output modes $b$ and 
$a$.}
\label{fig1}
\end{figure}

An entangled {\it retrodictive} state can be made as follows. Photon
detectors D1 and D0 are in the output ports {\it b} and {\it a }of a 50/50
beam splitter as shown in Fig. 1(b). If D1 detects one photon and D0 detects
zero photons, the output field of the beam splitter is in the retrodictive
state $\left| 1\right\rangle _b\left| 0\right\rangle _a$. This state
propagates backwards in time through the beam splitter. The unitary operator
for this evolution is $\hat{R}^{\dagger }$. We can show that 
\begin{equation}
\hat{R}^{\dagger }\left| 1\right\rangle _b\left| 0\right\rangle
_a=2^{-1/2}(\left| 1\right\rangle _b\left| 0\right\rangle _a-i\left|
0\right\rangle _b\left| 1\right\rangle _a)  \label{e6}
\end{equation}
which can be interpreted that the photon has an equal chance of being
reflected in its journey back in time into mode {\it a} or of staying in
mode {\it b}. It is sometimes convenient to regard D1, D0 and the beam
splitter as a combined measurement device that generates the retrodictive
state $2^{-1/2}(\left| 1\right\rangle _b\left| 0\right\rangle _a-i\left|
0\right\rangle _b\left| 1\right\rangle _a)$ corresponding to the combined
event of D1 detecting one photon and D0 detecting zero photons. If there is
a preparation device with a known output state acting on input mode {\it a},
then we can project this state onto the retrodictive entangled state,
resulting in a retrodictive state in input mode {\it b}. This retrodictive
state continues to evolve backwards in time.

Let us examine, in terms of the retrodictive formalism, the case where a
combined preparation device generates the predictive state $2^{-1/2}(\left|
0\right\rangle _b\left| 1\right\rangle _c+i\left| 1\right\rangle _b\left|
0\right\rangle _c)$ at time $t_p,$ for example as described above, and where
a later measurement on the field in mode {\it c }at time $t_m${\it \ }shows
the presence of one photon in this mode. We attach a retrodictive state $%
\left| 1\right\rangle _c$ to the field at the time of measurement. This
state evolves backwards in time and becomes $\hat{U}^{\dagger }\left|
1\right\rangle _c$ at the output of the combined preparation device, that is
at the output port of the beam splitter, at time $t_p$. Here $\hat{U}%
^{\dagger }$ is the hermitian conjugate of the free-space forward time
evolution operator from $t_p$ to $t_m.$ The effect of $\hat{U}^{\dagger }$on 
$\left| 1\right\rangle _c$ is to leave this state unchanged, so at the
preparation time $t_{p\text{ }}$we project $\left| 1\right\rangle _c$ onto
the state $2^{-1/2}(\left| 0\right\rangle _b\left| 1\right\rangle _c+i\left|
1\right\rangle _b\left| 0\right\rangle _c)$, leaving us after normalization
with the predictive state $\left| 0\right\rangle _b$ which evolves forward
in time. This is the same state that we obtained in the predictive
description of the same situation but there is an important difference. In
the retrodictive formalism the collapse to this state takes place at $t_{p%
\text{,}}$ the time of preparation, that is {\it as the field is leaving the
beam splitter}, whereas in the predictive formalism it occurs at $t_m,$ the
time of measurement, giving rise to the Einstein-Podolsky-Rosen paradox.

\section{UNCONTROLLABLE TIME MACHINE}

We can combine the above beam splitter with preparation devices P0 and P1 in
its input modes, which we shall refer to as BSL, and the above beam splitter
with photon detectors D0 and D1 in its output modes, which we shall refer to
as BSU, to form a double beam splitter arrangement that shares the common
mode {\it b}, as shown in Fig. 2 \cite{Sci}. BSL and BSU are the lower and
upper beam splitters in this figure. The output mode {\it b }of beam
splitter BSL becomes the input mode {\it b} of BSU. In the other input mode
of BSU, that is input mode {\it a}, we put a preparation device whose single
possible outcome event corresponds to the known predictive state 
\begin{equation}
\left| in\right\rangle _a=a_0\left| 0\right\rangle _a+a_1\left|
1\right\rangle _a  \label{e7}
\end{equation}
at the immediate entry to BSU. We assume also that we can adjust the
preparation device to control the ratio of the coefficients $a_0/a_1$. We
also know that P0 and P1 prepare vacuum and one-photon states respectively.
We can interpret the superposition state (\ref{e7}) as showing that one or
zero photons might be in input mode {\it a}. Thus, including the photon
input from P1, detectors D0 and D1 might detect a total of either one or two
photons. There is a sizeable probability that D0 and D1 will detect zero and
one photons respectively. We can calculate this probability, which depends
on $a_0/a_1,$ but we do not need the precise result here. Other possible
measurement outcomes are D0 and D1 detecting one and zero photons
respectively, one and one photon respectively, zero and two photons
respectively and two and zero photons respectively.

\begin{figure}
\includegraphics[width=0.30\textwidth]{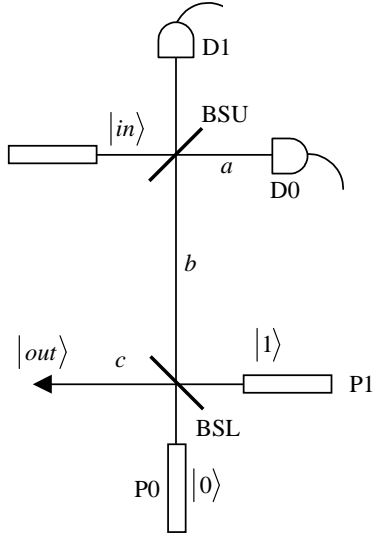}
\caption{Double beam splitter arrangement. The upper and lower beam
splitters BSU and BSL are an integer number of wavelengths apart. If
detectors D0 and D1 detect zero and one photons respectively, the state $
\left| out\right\rangle $ is the same as the state $\left| in\right\rangle $
.}
\label{fig2}
\end{figure}

Let us study the case in which detectors D0 and D1 detect zero and one
photons respectively. As seen in (\ref{e6}), this detection event generates
an entangled retrodictive state $2^{-1/2}(\left| 1\right\rangle _b\left|
0\right\rangle _a-i\left| 0\right\rangle _b\left| 1\right\rangle _a)$ at the
input to BSU. As we know there is a preparation event corresponding to a
predictive state $\left| in\right\rangle _a$ here, we project this state
onto the entangled state and obtain the retrodictive state that, after
normalization, is 
\begin{equation}
_a\left\langle in\right| (\left| 1\right\rangle _b\left| 0\right\rangle
_a-i\left| 0\right\rangle _b\left| 1\right\rangle _a)=a_0^{*}\left|
1\right\rangle _b-ia_1^{*}\left| 0\right\rangle _b.  \label{e8}
\end{equation}
This is the state of the field at the time $t_m$. This state propagates
backwards in time in mode {\it b} to become 
\[
\hat{U}^{\dagger }(a_0^{*}\left| 1\right\rangle _b-ia_1^{*}\left|
0\right\rangle _b) 
\]
at the output mode {\it b }port of the other beam splitter BSL at the
earlier time $t_p$. The action of the free space backward time evolution
operator $\hat{U}^{\dagger }$ is just to change the phase of the state $%
a_0^{*}\left| 1\right\rangle _b-ia_1^{*}\left| 0\right\rangle _b$. To avoid
unnecessary complications without losing the essential physics, we now
specify that the optical distance between the two beam splitters is an
integer number of wavelengths of the light. This means that the retrodictive
state at the earlier time $t_p$ at the output of BSL becomes again $%
a_0^{*}\left| 1\right\rangle _b-ia_1^{*}\left| 0\right\rangle _b$. At this
time this state is projected onto the known output $2^{-1/2}(\left|
0\right\rangle _b\left| 1\right\rangle _c+i\left| 1\right\rangle _b\left|
0\right\rangle _c)$ of the combined P0, P1 and BSL preparation device to
give 
\begin{equation}
2^{-1/2}(a_0\,_b\left\langle 1\right| +ia_1\,_b\left\langle 0\right|
)(\left| 0\right\rangle _b\left| 1\right\rangle _c+i\left| 1\right\rangle
_b\left| 0\right\rangle _c)=2^{-1/2}i(a_0\left| 0\right\rangle _c+a_1\left|
1\right\rangle _c).  \label{e9}
\end{equation}
After normalization and removal of the undetectable phase factor {\it i}, we
see that this predictive state in mode {\it c}, which we label $\left|
out\right\rangle _c$, is identical to the predictive state $\left|
in\right\rangle _a$ in (\ref{e7}) which we inject into BSU {\it at a later
time.}

We see therefore that, by means of the double beam splitter device, we can
send a state of light $\left| in\right\rangle =a_0\left| 0\right\rangle
+a_1\left| 1\right\rangle $, for which we can choose the ratio $a_0/a_1$,
backwards in time from when we prepare it at time $t_m$ to an earlier time $
t_p$. In principle the beam splitters can be separated by, say, one light
day so the state $\left| in\right\rangle $ which we choose today can be sent
back to appear yesterday. Can we use this device to send a message to our
earlier selves? Unfortunately, we cannot. The state $\left| in\right\rangle $
appears yesterday only if the detectors D0 and D1 detect zero and one
photons respectively and we have no control over what they will detect. If
they detect one and zero photons respectively, which is just as likely, then
the state $\left| out\right\rangle $ that appears yesterday is $a_0\left|
0\right\rangle -a_1\left| 1\right\rangle $. If they both detect zero photons
then $\left| out\right\rangle =\left| 1\right\rangle $. If they detect a
total of two photons then $\left| out\right\rangle =\left| 0\right\rangle $.
Thus in the absence of preknowledge of what will be detected the best we can
do is to send a mixed state back. This mixed state is the sum of the density
matrices representing the four possible states $\left| out\right\rangle $
weighted by the probabilities of them occurring. A formal calculation of
this mixed state and of the quantum information it can carry shows that zero
information can be transmitted to the past. This may seem a little
surprising because, if we could measure the state $\left| out\right\rangle ,$
sometimes we might find $\left| 1\right\rangle $ or $\left| 0\right\rangle $%
, which we could ignore, but on a sizeable number of occasions we would find 
$a_0\left| 0\right\rangle +a_1\left| 1\right\rangle $ or $a_0\left|
0\right\rangle -a_1\left| 1\right\rangle $ from which we could calculate the
ratio $a_0/a_1,$ which could be the number of the winner of a horse race.
Unfortunately, to determine the state $\left| out\right\rangle $ with any
reasonable precision, we need a large of identical copies of it. This cannot
be done by cloning and, as we have noted above, our lack of control over the
outcomes of D0 and D1 prevents us from sending identical copies back.

\section{A CLOSED CYCLE}

What use is an uncontrollable time machine if the lack of control is such as
to prevent us sending any information back in time? Even though we cannot
control it we do, however, send something back. Further, we do {\it know }%
what we have sent back. Thus it might be possible to use such a device to
examine a causal loop associated with a time travel paradox.

Specifically, let us consider the robot cycle described in the introduction.
The robot enters a time machine today in a particular state $S$ and is sent
back in time to yesterday. Upon emerging from the time machine yesterday it
enters a rejuvenator which adjusts its state and programs it to walk around
for a day and then to enter the time machine in precisely the state $S$.
Although the robot exists only during the cycle, the cycle is not isolated
from the rest of the world. As well as there being a need to build and
adjust the time machine and rejuvenator, the robot itself will leave
footprints as it enters the time machine. These footprints will provide
evidence of the existence of the robot as well as providing information
about its properties, for example its size. The action of the rejuvenator on
the robot may depend on properties of the robot itself. If the rejuvenator
fails to program the robot to be in the state $S$ today, either because of
bad adjustment or because of some property of the robot, then we have an
inconsistency. According to the principle of self consistency only
self-consistent loops are possible, so in this case there would be no
footprints associated with this particular type of robot entering the time
machine. The adjustment of the rejuvenator might, however, be suitable to
provide a self-consistent cycle for a different robot, say a small robot. In
this case, the need for consistency would not eliminate this cycle and any
footprints left would be small.

\begin{figure}
\includegraphics[width=0.40\textwidth]{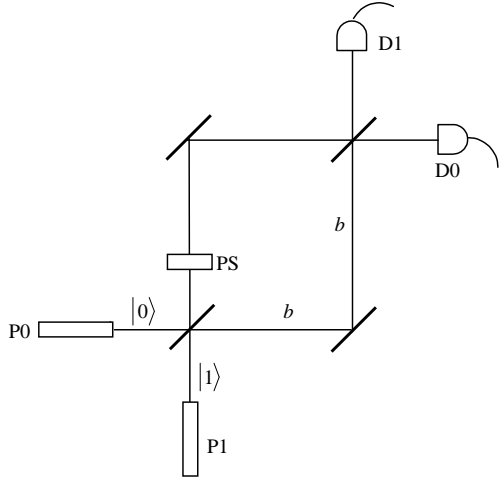}
\caption{Double beam splitter arrangement with feedback. The output light
in state $\left| out\right\rangle $ from the lower beam splitter is directed
through a phase shifter PS and via a fully reflecting mirror to become the
input light in state $\left| \overline{in}\right\rangle $ at the input of
the upper beam splitter.}
\label{fig3}
\end{figure}

Our aim is to use the double beam splitter device to examine closed cycles.
Classically if the rejuvenator programs the robot to be in a state today
even slightly different from $S$, an inconsistency arises because we regard
different classical states as effectively orthogonal. This is justifiable to
the extent that, in quantum mechanical terms, even if one particle of the
robot is in an orthogonal state to what it should be then the complete state
of the robot is orthogonal to what it should be. For simpler quantum
mechanical systems, however, the distinction is not as clear cut. Here we
have the possibility of complete self-consistency if the two states under
consideration are identical and of complete inconsistency if these states
are orthogonal. In between we have states that are not identical but not
orthogonal. We shall see the effect of this later. To create a consistent
cycle we wish to allow the field in state $\left| out\right\rangle $ at $t_p$
to evolve forward in time until, at $t_m$, it is in the state $\left|
in\right\rangle $ entering the input port of BSU in mode $a$. To do this we
create a feedback loop with fully reflecting mirrors while keeping the two
path lengths between BSL and BSU equal, as shown in Fig. 3. Again, for
simplicity, we assume both these path lengths are an integer number of
wavelengths. In order to experiment with consistent and inconsistent cycles
we also insert a phase shifter into the feedback part of the cycle. This
acts as our adjustable rejuvenator. Introducing the other part of the loop
in Fig. 3 removes the controllable preparation device which prepares $\left|
in\right\rangle $ in Fig. 2 so we no longer have the control we had to
choose the ratio $a_0/a_1$. The only preparation devices are now P0 and P1.
From conservation of energy, this means that a total of only one photon can
be detected by D0 and D1. Also, from symmetry, the photon should have an
equal probability of being found in mode {\it b} and in the mode containing
the phase shifter if we were to try to detect photons in these modes. That
is, there should be a probability of 1/2 of finding the photon in either
mode. We can ensure this by letting $\left| a_0\right| ^2=\left| a_1\right|
^2=1/2$.

The effect of applying a phase shift $\varphi $ to the field in state $%
\left| out\right\rangle $ is to change it from $a_0\left| 0\right\rangle
+a_1\left| 1\right\rangle $ to $a_0\left| 0\right\rangle +\exp (i\varphi
)a_1\left| 1\right\rangle ,$ so $\left| out\right\rangle $ will evolve to
the state 
\begin{equation}
\left| \overline{in}\right\rangle =a_0\left| 0\right\rangle +a_1\exp
(i\varphi )\left| 1\right\rangle  \label{e10}
\end{equation}
at time $t_m.$ Thus by adjusting the value of $\varphi $ to zero or an
integer multiple of $2\pi $, we can make the rejuvenator work perfectly and
obtain the consistency requirement $\left| \overline{in}\right\rangle =$ $%
\left| in\right\rangle $. For a general value of $\varphi $, we obtain the
projection 
\begin{equation}
\left\langle in|\overline{in}\right\rangle =\left| a_0\right| ^2+\exp
(i\varphi )\left| a_1\right| ^2=[1+\exp (i\varphi )]/2.  \label{e11}
\end{equation}
By setting $\varphi =\pi $ we see that (\ref{e11}) vanishes, so $\left| 
\overline{in}\right\rangle $ is {\it orthogonal} to $\left| in\right\rangle $
. In this case we can say that the rejuvenated state is definitely not the
state required for a self-consistent cycle. The principle of self
consistency would then imply that this closed cycle is impossible. The
outcome or ``footprints'' associated with this particular cycle, that is, D0
and D1 measuring zero and one photons respectively, should thus never be
observed.

Let us now keep $\varphi =\pi $ and examine the cycle we would obtain
associated with the other outcome that D0 and D1 measure one and zero
photons respectively. This is the only other possible outcome for the closed
cycle because, as noted above, a total of only one photon can be registered
by D0 and D1. We then find, in place of (\ref{e6}) that

\begin{equation}
\hat{R}^{\dagger }\left| 0\right\rangle _b\left| 1\right\rangle
_a=2^{-1/2}(\left| 0\right\rangle _b\left| 1\right\rangle _a-i\left|
1\right\rangle _b\left| 0\right\rangle _a).  \label{e12}
\end{equation}
We find that projecting $\left| in\right\rangle $ onto (\ref{e12}) gives us
the retrodictive state $-ia_0^{*}\left| 1\right\rangle _b+a_1^{*}\left|
0\right\rangle _b$ at $t_m.$ This state evolves backwards in time to $%
-ia_0^{*}\left| 1\right\rangle _b+a_1^{*}\left| 0\right\rangle _b$ at $t_p$
because the mode {\it b} path is an integer number of wavelengths.
Projecting this onto the prepared entangled state $2^{-1/2}(\left|
0\right\rangle _b\left| 1\right\rangle _c+i\left| 1\right\rangle _b\left|
0\right\rangle _c)$ gives, after normalization and removal of the
undetectable factor, the state 
\begin{equation}
\left| out\right\rangle =a_0\left| 0\right\rangle _c-a_1\left|
1\right\rangle _c.  \label{e13}
\end{equation}
This leads to 
\begin{equation}
\left| \overline{in}\right\rangle =a_0\left| 0\right\rangle -a_1\exp
(i\varphi )\left| 1\right\rangle =a_0\left| 0\right\rangle +a_1\left|
1\right\rangle .  \label{e14}
\end{equation}
We see that for this case we do have a self-consistent cycle, with
associated ``footprints'' being D0 and D1 measuring one and zero photons
respectively. Thus there is no reason on the basis of the principle of self
consistency to eliminate this cycle. On the other hand if we set $\varphi =0$
we find for this case that $\left| \overline{in}\right\rangle =a_0\left|
0\right\rangle -a_1\left| 1\right\rangle $ so we have an inconsistent cycle
so there should be no chance of D0 and D1 measuring one and zero photons
respectively for this setting of $\varphi $.

We have seen that if we use the only output measure we have, that is what D0
and D1 detect, as a ``probability meter'' for a cycle occurring, then the
principle of self consistency implies that we can adjust the setting of $
\varphi $ to make one or other of the two possible measurement outcomes
impossible. While this is really all we can talk about classically, in
quantum mechanics we also have intermediate cases. For example, what happens
if we choose $0<\varphi <\pi $? In this case there would be some incomplete
overlap of $\left| in\right\rangle $ and $\left| \overline{in}\right\rangle
. $ We might expect that we could extend the principle of self consistency
such that the probability for the cycle occurring is$\left| \left\langle in|
\overline{in}\right\rangle \right| ^2$. Thus on repeating the experiment a
large number of times, in some of cases the footprints for one cycle would
be found, for example D0 and D1 measuring one and zero photons respectively,
and for the remaining cases the alternative footprints would be found.

\section{INTERFERING AMPLITUDES}

The next question is whether or not we should build the uncontrollable time
machine with the feedback loop to investigate experimentally the principle
of self consistency, that is, to check if we do indeed find it impossible
for D0 and D1 to detect zero and one photons respectively if $\varphi =\pi $
. Fortunately there is no need actually to do the experiment because what we
have shown in Fig. 3 is essentially a Mach-Zehnder interferometer for which
we know the results. As shown below, these confirm that D0 and D1 do not
detect zero and one photons respectively if $\varphi =\pi $ and do not
detect one and zero photons respectively if $\varphi =0$. Even more
fortunately, we can explain how the Mach-Zehnder interferometer works in
terms of interference of quantum mechanical amplitudes in the usual
predictive formalism of quantum mechanics. Essentially there are two paths
by which a photon from P1 in Fig. 3 can be detected by D1, that is, via the
mode {\it b} path and via the phase shifter path. These two paths are the
two components of the cycle. We might think of the photon being reflected by
BSL and then transmitted by BSU or as being transmitted by BSL and reflected
by BSU. The amplitude for reflection $A(r)$ by BSL into mode {\it b} is the
coefficient of $\left| 1\right\rangle _b\left| 0\right\rangle _c$ in (\ref
{e5}), that is $A(r)=i2^{-1/2}.$ Likewise the transmission amplitude for BSU
is the coefficient of $\left| 0\right\rangle _b\left| 1\right\rangle _c$ on (
\ref{e5}), that is, $A(t)=2^{-1/2}.$ Thus the compound amplitude for
reaching D1 via mode {\it b} is $A(r)A(t)=i/2$. The transmission
amplitude for BSL is also $A(t)$ and phase shifter introduces another factor 
$\exp (i\varphi ),$ so the amplitude of being transmitted and reaching BSU
is $A(t,\varphi )=2^{-1/2}\exp (i\varphi )$. The reflection amplitude for
BSU is also $A(r).$ The compound amplitude for reaching D1 via the mode
containing the phase shifter is $A(t,\varphi )A(r)$. The {\it total}
amplitude for detection of the photon by D1 is then 
\begin{equation}
A(r)A(t)+A(t,\varphi )A(r)=i[1+\exp (i\varphi )]/2.  \label{e15}
\end{equation}

If $\varphi =\pi $ the amplitude for one path is the negative of the
amplitude for the other path and the total amplitude, and therefore the
probability for D1 to detect a photon, is zero. If $\varphi =0,$ however,
the amplitudes for these two paths add constructively, giving a unit
probability for detection. We note that the probability obtained by squaring
the modulus of (\ref{e15}) is equal to$\left| \left\langle in|\overline{in}%
\right\rangle \right| ^2.$ These results confirm the principle of self
consistency for this particular case {\it and} its extension that the
probability for footprints associated with the cycle occurring is$\left|
\left\langle in|\overline{in}\right\rangle \right| ^2$.

These results indicate that we should look to quantum mechanical {\it %
amplitudes}, rather than probabilities, in seeking the physical mechanism
underlying the impossibility of inconsistent cycles, that is, in
understanding how physics sorts things out and arrives at a consistent
cycle. For a causal closed cycle, we can choose the earliest event and the
latest event, which is associated with the footprints for example, and then
regard the two parts of the cycle as two different ways of reaching the
latest event, or footprints, from the earliest event. Indeed, we could also
regard the cycle as two different ways of reaching the earliest event from
the latest event. If the amplitudes associated with the two parts completely
interfere then the total amplitude associated with the cycle is zero and the
probability for the cycle occurring is zero, that is, it is impossible.
Constructive interference, however, renders the cycle possible. For simple
quantum systems, there are intermediate cases between completely destructive
and completely constructive interference, so the probability for some cycles
will be somewhere between zero and unity. There only needs to be a slight
difference in two states of a macroscopic object, for example if the two
states are identical except for the states of just one particle of the
object which are orthogonal, to make the macroscopic states orthogonal
themselves. Consequently we usually consider classical cycles as being
either consistent or inconsistent.

\section{CONCLUSION}

The possibility of inconsistent closed causal cycles has been used as an
argument against the possibility of backward time travel$.$ Against this,
the principle of self consistency has been proposed which states that
physics sorts things out so that inconsistent cycles are impossible anyway,
so any closed cycles must be consistent. If this principle is correct then
the possibility of closed cycles is not a valid argument against backward
time travel. This principle, however, opens up another question - how does
physics sort out things so that only consistent cycles occur? In this paper
we have examined this question in the light of a device which can be
interpreted, in terms of the retrodictive formalism of quantum mechanics, as
an uncontrollable time machine. A known state of light $\left|
in\right\rangle $ at the input of a beam splitter at time $t_m$ is sent into
the past and then allowed to evolve into the future so that at time $t_m$ it
is in state $\left| \overline{in}\right\rangle $ at the input of the beam
splitter. Because the device is interpretable in usual predictive quantum
theory as an interferometer, we can calculate the probability of the cycle
occurring and leaving its particular footprints to be $\left| \left\langle 
\overline{in}|in\right\rangle \right| ^2$. This means that the probability
of the cycle occurring, as observed by its footprints, is unity if $\left| 
\overline{in}\right\rangle $ is the same state as $\left| in\right\rangle $
and is zero if $\left| \overline{in}\right\rangle $ is orthogonal to $\left|
in\right\rangle $. The former case is a self consistent cycle and the latter
is an inconsistent cycle. These results {\it confirm} the principle of self
consistency for this particular case. Furthermore, the results allow a
possible extension of the principle, which is essentially classical, to
quantum mechanics to say that the probability of a cycle occurring can be
found by calculating the quantum evolution of the state around the cycle,
which involves evolution both backwards and forwards in time, to find the
evolved state at the starting point. The probability of the cycle is then
given by the square of the modulus of the projection of the evolved state
onto the original state.

If what we have found for the cycle we have studied can be applied to all
cycles, then we have an underlying quantum mechanical explanation of the
principle of self consistency. Essentially, if we work in terms of quantum
mechanical amplitudes rather than probabilities, we can select the latest
and earliest events on the cycle and then say that the amplitude for
reaching the latest event from the earliest event has two terms,
corresponding to the two different pathways to the later event. For
inconsistent cycles these two amplitudes cancel each other when added. For
consistent cycles, they interfere constructively. The probability of the
cycle is the square of the modulus of the total amplitude and thus fully
inconsistent cycles have a probability of zero of occurring, that is, they
are impossible. Even slightly different classical states, that is quantum
states of macroscopic objects, can be orthogonal so these slight differences
induced by evolution around the cycle can render the cycle impossible.
Classical cycles are thus usually regarded as consistent or inconsistent.

It is not totally surprising that a principle applying to classical physics
has a quantum mechanical basis. The classical principle of least action can
be explained in terms of the addition of amplitudes associated with all
possible paths. The amplitudes for all paths except for those in the region
of the path of least action cancel, so the probability for finding that the
system has taken a path not near the path of least action is zero \cite{Fey}%
. This explains how the system ``knows'' to take the path of least action.
In this paper we suggest that closed causal cycles are sorted out by a
similar mechanism. Only those cycles with a net non-zero amplitude have a
non-zero probability of occurring and these are the consistent cycles. In
conclusion, rather than just being invoked to save the possibility of the
present shaping the past, it now seems that the principle of self
consistency could well have a solid physical basis in quantum mechanics.

\section*{Acknowledgments}

I thank S. Barnett, J. Jeffers, O. Jedrkiewicz and R. Loudon for discussions
on retrodiction. This work was supported by a grant from the Australian
Research Council.


\begin{thebibliography}{99}
\bibitem{Nap}  D. T. Pegg, in {\it Time's Arrows, Quantum Measurement 
and Superluminal Behavior} edited by D. Mugnai, A. Ranfagni and L. S. 
Schulman (Consiglio Nazionale Delle Richerche, Roma, 2001) p. 113.
    
\bibitem{Nov}  See for example I. D. Novikov, {\it The River of Time}
(Cambridge University Press, Cambridge, 1998).

\bibitem{Hor}  P. Horwich, in {\it Time's Arrows Today} edited by S. F.
Savitt (Cambridge University Press, Cambridge, 1997).

\bibitem{Sch}  J. A. Wheeler and R. P. Feynman, Rev. Mod. Phys. 21, 425
(1949); A. Peres and L. S. Schulman, Phys. Rev. D 5, 2654 (1971); L. S.
Schulman, J. Math. Phys. 15, 296 (1974).

\bibitem{Ear}  J. Earman, in {\it Time's Arrows Today} edited by S. F.
Savitt (Cambridge University Press, Cambridge, 1997).

\bibitem{Feyn}  I have heard this scenario attributed to Feynman but do not
know the reference.

\bibitem{He}  C. W. Helstrom, {\it Quantum Detection and Estimation Theory}
(Academic Press, NewYork, 1976).

\bibitem{At}  S. M. Barnett, D. T. Pegg, J. Jeffers and O. Jedrkiewicz , J.
Phys. B: At. Mol. Opt. Phys.33, 3047 (2000).

\bibitem{Bay}  S. M. Barnett, D. T. Pegg and J. Jeffers, J. Mod. Opt., 47,
1779 (2000).

\bibitem{Ah}  Y. Aharonov, P. G. Bergman and J. L. Lebowitz, Phys. Rev. 134,
B1410 (1964); R. H. Penfield, Am. J. Phys. 34, 422 (1966); Y. Aharonov and
D. Z. Albert, Phys. Rev. D 29, 223 (1984); Y. Aharonov and D. Z. Albert,
Phys. Rev. D 29, 228 (1984); Y. Aharonov and L. Vaidman, J. Phys. A: Math.
Gen. 24, 2315 (1991).

\bibitem{Amp}  S. M. Barnett, D. T. Pegg, J. Jeffers, O. Jedrkiewicz and R.
Loudon Phys. Rev. A 62, 022313 (2000).

\bibitem{Book}  See for example S. M. Barnett and P. M. Radmore Methods in
Theoretical Quantum Optics (Oxford University Press, Oxford, 1997).

\bibitem{Sci}  D. T. Pegg, L. S. Phillips and S. M. Barnett, Phys. Rev.
Lett. 81 1604 (1998); D. T. Pegg and S. M. Barnett, Quantum and Semiclass.
Opt. 1, (1999); S. M. Barnett and D. T. Pegg, Phys. Rev. A 60, 4965 (1999).

\bibitem{Fey}  R. P. Feynman, R. B. Leighton and M. Sands, {\it The Feynman
Lectures in Physics }Vol. II (Addison-Wesley, Reading, 1966) p. 19-9.
\end{thebibliography}
\end{document}